\documentclass[a4paper]{article}

\usepackage{graphicx}
\usepackage{subfig}
\usepackage{natbib}

\begin{document}

\begin{center}
\begin{Large}
{\bf Recent Results of Solid-State Spectroscopy}
\end{Large}
\end{center}

\vspace{0.8cm}

\noindent{}Cornelia J\"ager$^{1,2}$, Thomas Posch$^3$, Harald Mutschke$^1$, 
Simon Zeidler$^1$, Akemi Tamanai$^4$, Bernard L.\ de Vries$^5$ \\

\noindent{}$^1$Astrophysikalisches Institut,
Friedrich-Schiller-Universit\"at Jena, Schillerg\"asschen 2-3, D-07745 Jena, Germany \\
$^2$Laboratory Astrophysics and Cluster Physics Group, Max Planck Institute for
Astronomy and Institute of Solid State Physics, Friedrich-Schiller-Universit\"at Jena, Max-Wien-Platz 1, D-07745 Jena, Germany,\\
Email: {\tt cornelia.jaeger@uni-jena.de} \\
$^3$Institut f\"ur Astronomie, T\"urkenschanzstra{\ss}e 17,
A-1180 Vienna, Austria \\
$^4$Kirchhoff-Institut f\"ur Physik, Universit\"at Heidelberg,
Im Neuenheimer Feld 227, D-69120 Heidelberg, Germany \\
$^5$Instituut voor Sterrenkunde, Katholieke Universiteit Leuven, Celestijnenlaan 200D, 3001 Leuven, Belgium \\

\noindent{}\textit{To be published in: 
Proceedings of the IAU Symp.\ 280, ``The Molecular Universe'',
p.\ 416-430}. \textit{Editors: J.\ Cernicharo \& R.\ Bachiller}\\



\noindent{}{\bf Abstract:}\\
Solid state spectroscopy continues to be an important source of
information on the mineralogical composition and physical properties
of dust grains both in space and on planetary surfaces. With only a
few exceptions, artificially produced or natural terrestrial analog
materials, rather than 'real' cosmic dust grains, are the subject of
solid state astrophysics. The Jena laboratory has provided a large
number of data sets characterizing the UV, optical and infrared
properties of such cosmic dust analogs. The present paper highlights
recent developments and results achieved in this context,
focussing on 'non-standard conditions' such as very low
temperatures, very high temperatures and very long wavelengths.\\

\noindent{}{\bf Keywords:}\\
{Methods: laboratory; ISM: dust, extinction; stars: AGB and
post-AGB; circumstellar matter}


\section{Introduction}

Solids are strangers in space. By far the largest part of the
baryonic matter in the universe is in the gas or plasma state. Less
than 1\% of cosmic matter consists of solids, such as ices, dust
grains or terrestrial planets. Yet, many of the molecules known from
astrochemical investigations would not have formed without the
catalytic role that the surfaces of solids play in various
environments (e.g.\ in circumstellar shells, in the interstellar medium,
protoplanetary disks, etc.).

Proper understanding of solids in space requires studies on analog
materials in terrestrial laboratories. Here we shall focus on the
{\em infrared spectra}\/ of cosmic dust analogs (CDAs), which are a
small, but crucial subset of CDA properties. An obvious goal of
infrared spectroscopy of CDAs is to create databases of dielectric
functions enabling us to identify cosmic dust species from their
observed astronomical spectra.
One such database is the Jena Database of Optical Constants of Solids:\\
{\tt http://www.astro.uni-jena.de/Laboratory/Database/databases.html}.\\
For most of the solids discussed here, data files of their
optical constants can be retrieved from this website. It should be
noted that while a precise analytical characterization has been
performed for all materials that are present in the Jena Database,
the respective detailed results of physico-chemical analyses are not
contained there, but only in the pertinent scientific articles.

Our paper is structured as follows. Section 2 is on mid-infrared
spectroscopy of solids, while section 3 is on far-infrared
spectroscopy. In both sections, we focus on recent measurements, a
substantial part of which were performed at either cryogenic
temperatures (especially in the case of the FIR-measurements) or at
high temperatures, namely in order to explore conditions under which
cosmic dust doubtlessly exists. At some points, a comparison to
astronomical infrared spectra is made, as a part of the above
mentioned, still ongoing effort to identify the main components of
circumstellar, interstellar and in protoplanetary dust.


\section{Mid-Infrared Spectroscopy}


\subsection{General considerations}

As for the identification of cosmic dust species on the basis of
their vibrational bands, the mid-infrared region (5-50\,$\mu$m) turned out to be
the most fruitful wavelength range.

For some species of solids -- e.g.\ crystalline silicates and highly
ionic crystalline oxides in which intense dipole oscillations can be
excited -- MIR powder spectra are strongly influenced by the shapes,
sizes, and possible agglomeration of the respective particles (e.g.\
Salisbury et al.\ 1992). Crystalline grains which are \textit{not}\/
small compared to the examined wavelengths, as well as powders
containing grains strongly deviating from spherical symmetry, show
markedly different spectra (band positions, band profiles) than very
small spherical grains. For amorphous solids, the influence of the
particle shapes on the infrared (powder) {spectra is} less
pronounced.

There are several ways to overcome or 'control' the
grain-shape-dependence of infrared spectra. One method is to
determine optical constants, $n$ and $k$ -- e.g.\ from reflectance
measurements of polished surfaces of bulk material -- and to use
these quantities for the calculation of small particle spectra for
arbitrary grain shapes (e.g.\ Bohren \& Huffman 1983). Compared to
mineral powder spectra -- for which the actual grain shapes are
often unknown or difficult to disentangle --, calculated small
particle spectra are better suited for a comparison with the spectra
of cosmic dust, since the grain shape can then be varied in the
calculations in a controlled way. Another method to get grip on
grain-shape-dependence is aerosol spectroscopy; exemplary results
are presented in the following subsection.


\subsection{Effects of grain shapes on band profiles examined by aerosol spectroscopy}

The Jena laboratory astrophysics group has adopted the aerosol
technique and performed IR spectroscopic measurements in a
matrix-free environment (Tamanai et al.\ 2006, 2009a, 2009b). With
this method, it is possible to directly study the influence of
particle morphologies on IR spectra. Any {spectral} influence of
a matrix material used in powder spectroscopy (e.g. potassium
bromide, KBr) is ruled out in aerosol spectroscopy, simply because
there is no matrix substance in this case (the particles are free
floating).

\begin{figure}
    \centering
        \subfloat[]{\includegraphics[width=1.5in]{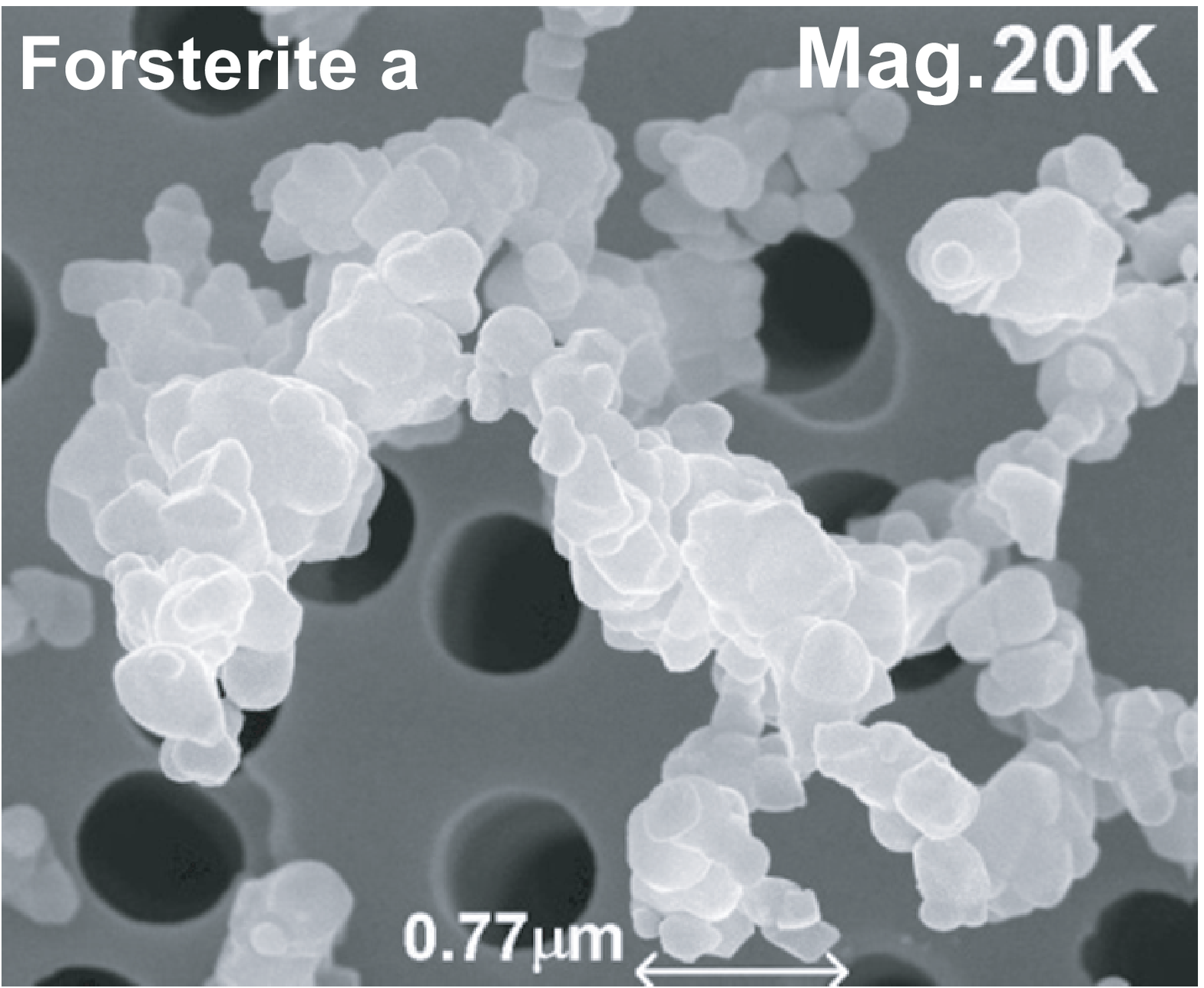}}
        \hspace{1cm}
        \subfloat[]{\includegraphics[width=1.5in]{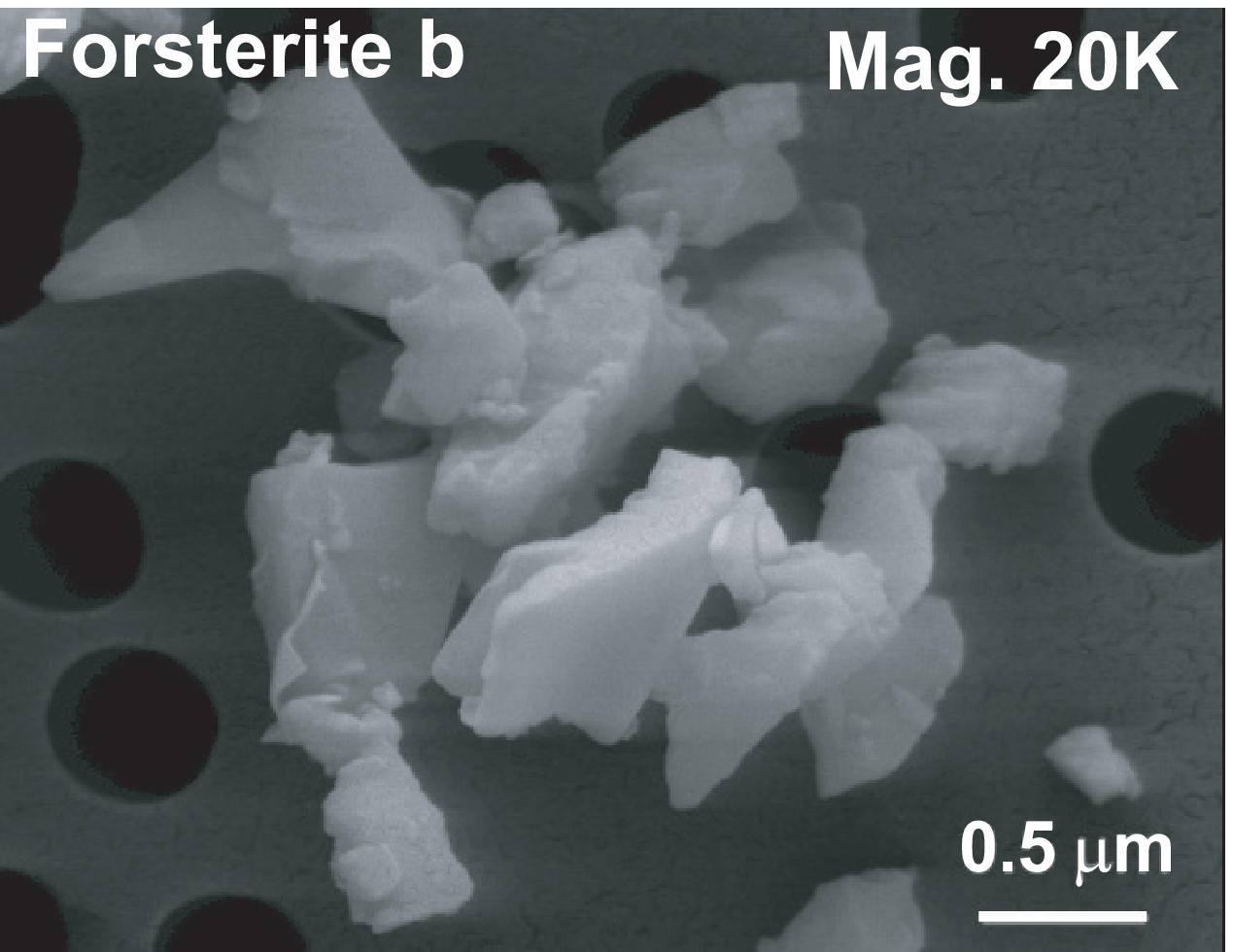}}
        \caption{SEM images of forsterite particles. (a) 'Forsterite a', with
        nearly spheroidal and ellipsoidal (rounded) grain shapes. (b) Irregular
        shapes with sharp edged particles dominating in 'Forsterite b'. Note: the
        black dots are the holes of the polyester membrane filter which have a diameter of ~0.5\,$\mu$m.}
    \label{ForsGrain}
\end{figure}

\begin{figure}
    \centering
        \includegraphics[width=2.9in]{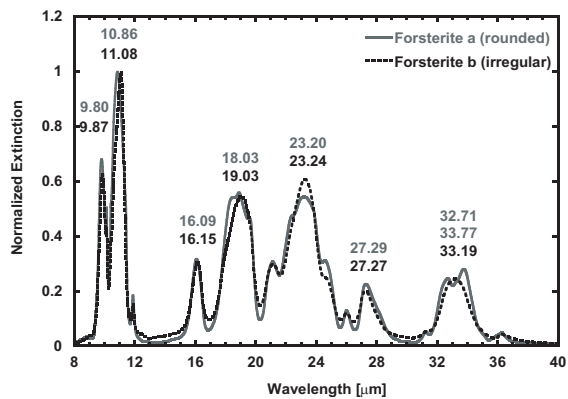}
    \caption{Normalized extinction vs.\ wavelength for two different shapes of 
    forsterite
    particles. The grey solid line refers to the 'Forsterite a' (roundish 
    grains), while the black dotted
    line refers to the 'Forsterite b' (irregular shapes).}
    \label{AeroSpec}
\end{figure}

Images of two
forsterite (Mg$_2$SiO$_4$) samples with different shapes obtained by
a scanning electron microscope (SEM) are shown in Fig.\
\ref{ForsGrain}. The image of 'Forsterite a' (left) is dominated by
roundish grains. On the other hand, 'Forsterite b' particles (right)
are irregularly shaped and show sharp edges. X-ray diffraction
has been used to examine the crystal structure of both samples,
which turned out to be  identical. As shown in Fig.\ \ref{AeroSpec},
the band profiles measured for forsterite particles with ellipsoidal
(roundish) shapes {clearly differ from those} with shapes
including sharp edges. The peaks of the resonances undergo a
blueshift of up to 0.22\,$\mu$m for the rounded grains (Tamanai et
al.\ 2006). Although it is difficult to demonstrate that the peak
shift is purely caused by the grain shape, a blueshift of the
rounded grains is expected, since the resonances of a sphere arise
at shorter wavelengths, which can be predicted by means
of theoretical simulations (Bohren \& Huffman 1983). It can be
predicted that resonances of particles far from a spherical shape,
such as needle-like ones, are located predominantly at the
transverse optical (TO) frequency, which is located at longer
wavelengths than the resonance of spherical grains. Another
interesting difference between the IR band profiles of rounded
and of irregular shaped particles is that rounded grains may produce
double peaked (rectangular) profiles (see e.g.\ the peaks at 19, 23,
33\,$\mu$m). This tendency has been observed in oxide particle
investigations as well (see more details in Tamanai et al.\ 2009a,
b).

The particle size and agglomeration state may partially affect the
spectral features as well. Thus it is essential to do
theoretical modeling in order to distinguish grain shape effects
from grain size and agglomeration effects.

\subsection{High temperature measurements of spinel and corundum and comparison with the observed 'astronomical' 13\,$\mu$m band}

A second 'hot topic' in solid-state astrophysics is to fully cover
the range of temperatures that dust grains reach in space. In the case
of circumstellar dust, this may imply the necessity to heat the solid
samples up to 1000\,K.

\begin{figure}
\vspace*{-0.7 cm}
\begin{center}
\subfloat[]{\includegraphics[width=2.5in]{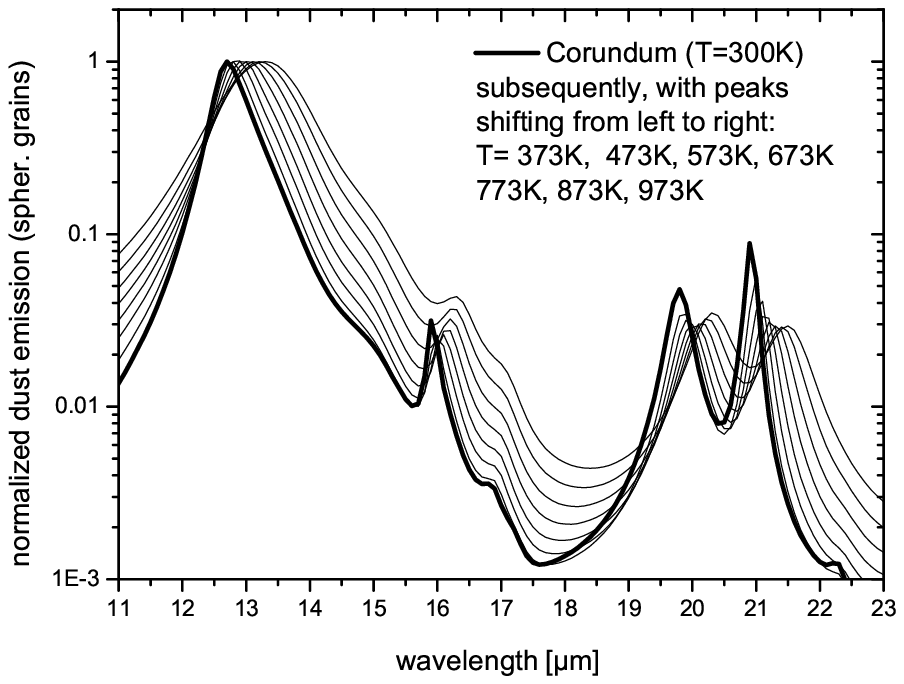}}
\subfloat[]{\includegraphics[width=2.5in]{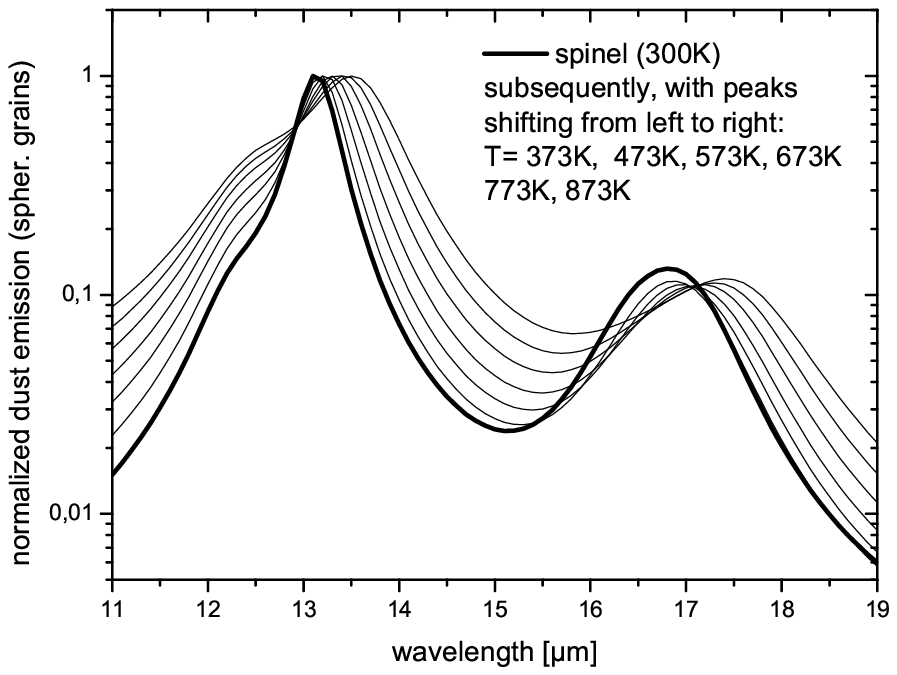}}
\vspace*{-0.3 cm}
 \caption{Normalized emission spectra of small spherical corundum (a) and
spinel grains (b) for different temperatures ranging from 300 to 973\,K.
Orientational averaging has been performed
as required by the anisotropy of $\alpha$-Al$_2$O$_3$. Note the shift of the peaks to
longer wavelengths and the increasing bandwidth as the temperature increases.}
   \label{corT}
\end{center}
\end{figure}

Two different cases of high temperature measurements have to be distinguished:

\begin{itemize}

\item{\textit{Annealing}}: In this case a sample undergoes some persistent structural transition
upon heating and keeping it hot for a certain time. The effect of the respective
structural change on the dielectric properties, crystal structure
etc.\ can be measured after cooling the sample to room temperature.

\item{\textit{'In situ'} high-T-measurements}: These have to be applied if
reversible changes (of infrared properties etc.) are expected to occur for
a cosmic dust analog, such that for instance its IR reflectance continuously
changes upon heating. IR spectra
consequently have to be taken while the sample is still hot, which may require
sophisticated experimental setups.

\end{itemize}

While annealing experiments have been carried out for cosmic dust
analogs in a number of laboratories, in situ high-T-spectra are so
far available for a few substances only.

We performed in situ high-T measurements of the IR properties of
corundum ($\alpha$-Al$_2$O$_3$) and spinel (MgAl$_2$O$_4$), since
{these dust species} are indeed expected to condense
at high temperatures (above 1000\,K) in stellar envelopes (see Gail
2010) and protoplanetary disks.

Both mineral species have been discussed as potential carriers of
the so-called 13$\mu$m band -- a noteworthy infrared feature
detected in oxygen-rich AGB stars with low mass loss rates (Posch et
al.\ 1999, Fabian et al.\ 2001, Sloan et al.\ 2003). Our measurements
set new constraints on the carrier substance of this band
and on its possible range of temperatures, as will be discussed below.

For our laboratory measurements, a high-temperature high-pressure
(HTHP) cell was used. This water-cooled cell allowed us to carry out
the temperature-dependent measurements up to 1073\,K. The HTHP cell
(type Specac P/N 5850) was integrated into a Bruker 113v
Fourier-Transform-Infrared spectrometer. {We measured reflection
spectra of crystallographically oriented minerals heated to 573, 773,
and 973 K, relative to the reflectance
of a gold mirror (kept at room temperature).} From the reflectance
spectra, the complex refractive indices of the minerals have been
calculated using a Lorentz oscillator model. In the case of spinel,
we used a Lorentz model with five oscillators, while for corundum,
11 and 9 oscillators for the crystallographic orientations
E$\parallel$c and E$\perp$c were needed, respectively. With
increasing temperature, the oscillator frequencies generally shift
to longer wavelengths and the {damping parameters increase}
strongly, while the oscillator strength changes only slightly. This
behavior is expected for resonances in an anharmonic vibrational
potential (e.g.\ Henning \& Mutschke 1997).

\begin{figure}
\begin{center}
 \includegraphics[width=3.4in]{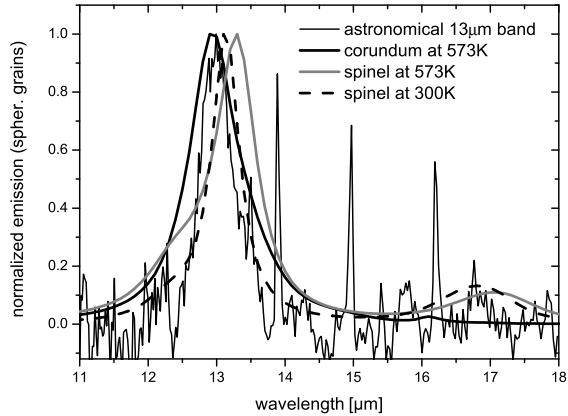}
\vspace*{-0.5 cm}
 \caption{Comparison between the average profile of the astronomical
 13$\mu$m emission band -- as derived by Fabian et al.\ (2001) for a
 sample of oxygen-rich AGB stars -- with the normalized emission
 spectra of corundum and spinel grains
at temperatures of {300\,K} and 573\,K.}
   \label{13mic}
\end{center}
\end{figure}

In order to derive the complex refractive indices ($n$+i\,$k$) --
also for intermediate temperatures -- we fitted the temperature
dependence of the oscillator parameters by second order polynomials.
In consequence, we obtained analytical expressions for the
temperature-dependent optical constants of the measured mineral for
each crystallographic axis.

From the optical constants, small-particle spectra
$Q$($\lambda$) were calculated for the simplest case of spherical
grains in the Rayleigh size limit (particle size $a$ $\ll$
$\lambda$). The emitted flux at a given temperature T finally
results from multiplying $Q$($\lambda$) with the corresponding
T-dependent blackbody function $B_{\nu}$($T$). The respective
emission spectra for temperatures ranging from 300\,K to 973\,K are
shown in Fig.\ \ref{corT}.

In Fig.\ \ref{13mic}, we compare the small-spherical-grain
{emission spectra}
of spinel and corundum to the average profile of the above mentioned
13\,$\mu$m band as derived by Fabian et al.\ (2001).

The comparison shows that \\
i) both spinel and corundum grains -- if mainly radiating at a
'suitable' narrow range of temperatures -- can account for the average
profile of the 13\,$\mu$m band: spinel at about 300\,K and corundum at about 573\,K; \\
ii) at high grain temperatures, the resulting integrated emission
both of spinel and of corundum grains becomes too broad as to fit
the observed astronomical band (see esp.\ the spinel spectrum for
573\,K; cf.\ Fig.\ 3a for corundum at T $>$ 600\,K).

As for corundum, it should be noted that only our high temperature
measurements enable a successful fitting to the 13\,$\mu$m band.
However, the problem remains that corundum has additional weaker
bands around 16\,$\mu$m and 20--21\,$\mu$m which have no counterparts in
the observed astronomical spectra.


\subsection{Spectral properties of carbonaceous material}

We now turn to the spectral properties of carbonaceous ma-
terials -- another major component of cosmic dust -- and start this subsection with a brief introduction into the secrets of carbonaceous
structures prepared by gas-phase condensation and with a short
overview on possible formation processes of carbonaceous solids
in astrophysical environments.
The structure of particulate carbon materials can be characterized
on different length scales describing the short-, medium-, and
long-range order of the material. The long-range order describes the
size and shape of the primary particles and their agglomeration
state. The medium-range order characterizes the arrangement, shape
and size of the structural subunits and can be derived by analytical
characterization such as high-resolution transmission electron
microscopy. The hybridization state of carbon, the nature of the
bonds between carbon atoms as well as the incorporation of hydrogen
or heteroatoms in the carbon grains can be specified by the
short-range order of the material (\cite{Schnai:97,Jaeger:99}). The
short- and medium-range order of carbon grains can be well
described by the terms 'amorphous carbon' (AC) or 'hydrogenated
amorphous carbon' (HAC). HACs are frequently used to describe the
carbonaceous dust component in astrophysical studies
(Jones et al.\ 1990, Duley 1994, Pendleton \& Allamandola 2002,
Dartois 2011). In AC and HAC, the carbonaceous structure can be
characterized by aromatic islands of different sizes, which are
linked by aliphatic structures including sp, sp$^2$, and sp$^3$
carbon atoms. In the astrophysical literature, the term HAC is
preferentially employed for structures containing aromatic units of
2-8 rings (Jones et al.\ 1990).

In recent laboratory studies on the gas-phase condensation of
carbonaceous material, a low- and a high-temperature formation
process has been found (\cite{Jaeger:09}). At condensation
temperatures higher than 3500\,K, fullerene fragments or complete
fullerenes build up the nucleating particles. Fullerene-like carbon
grains and fullerene compounds are formed. Low-temperature condensation
(T$\leq$1700\,K) favours a nucleation and growth process with PAHs as
precursors and particle-forming elements resulting in carbonaceous
grains revealing well developed planar or slightly bent graphene
layers in their interior and a mixture of PAHs as side products.

\begin{figure}
 \vspace*{-1.0 cm}
\begin{center}
 \includegraphics[width=3.4in]{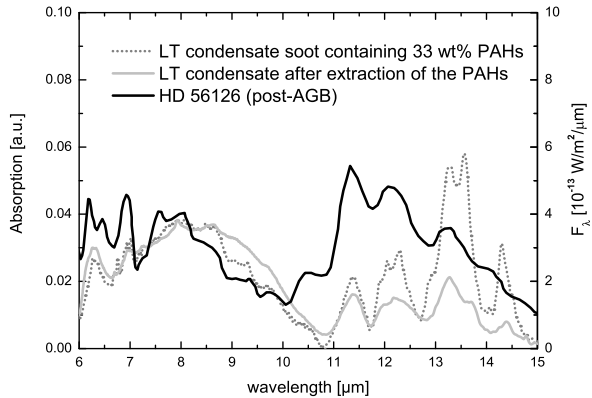}
 \vspace*{-0.3 cm}
 \caption{Comparison between the IR
spectral properties of a carbonaceous condensate that consists of a
mixture of solid carbon particles and PAHs and the observed spectrum
of a post AGB star adapted from Hony et al.\ (2003).}
   \label{carbIR}
\end{center}
\end{figure}

The MIR spectra of gas-phase condensed carbonaceous solids that may
consist either of a mixture of grains and soluble components (PAHs
or fullerenes) or of carbonaceous grains exclusively, without a
soluble component, show aromatic IR bands (AIBs) as well as
aliphatic IR band, partly superimposed on broad plateau features
around 8 and 12\,$\mu$m (see Fig.\,\ref{carbIR}).
Aliphatic --CH$_x$ absorptions can be observed at 3.4, 6.8, and
7.25\,$\mu$m. In fullerene-like carbonaceous grains, saturated
aliphatic --CH$_x$ groups are mainly responsible for the links
between the fullerene fragments (\cite{Jaeger:08b}). In this
material, signatures related to the presence of --C$\equiv$C-- triple bonds
can be observed in the in-situ recorded IR spectra at 3.0 and
4.7\,$\mu$m, which points to the presence of polyynes. For the
formation of fullerene-like soot grains, polyynes are discussed as
intermediates. They finally form up the fullerene snatches
that are incorporated into the grains. Additional bands due to
--C--C-- stretching, --C--H, and --C--C-- twisting, wagging, and
rocking bands can be expected in the range between 7.4 and
10\,$\mu$m. In the 11-15\,$\mu$m range, out-of-plane --C--H bending
bands can be observed caused by --C--H groups at the periphery of the
fullerene snatches.

The more aromatic carbonaceous material produced by low-temperature
condensation of hydrocarbons in laser pyrolysis (\cite{Jaeger:09})
show a series of aromatic bands including 3.3, 6.3, 8.6, 11.3, 12.3,
and 13.3\,$\mu$m as well as aliphatic IR bands at 3.4, 6.8, and
7.25\,$\mu$m (--C--H stretching and deformation bands). IR features
observed beyond 6.7\,$\mu$m (6.8, 7.25, 7.93, 8.47, 9.28, 9.72, 10.5
\,$\mu$m) mainly arise from --C--C-- stretching, C-H deformation,
and other bands due to --C--H twisting, wagging, rocking, and
--C--C-- deformation bands. In this spectral range, a firm assignment
of the IR features to a specific functional group is rather
difficult even for carbonaceous grains without adsorbed PAHs, which
may contain a lot of different functional groups with single and
double bonds integrated in varying chemical environments leading to
a distribution of vibrational bands in this spectral range.

Bands of more diagnostic relevance can be found at wavelengths
larger than 11\,$\mu$m, which can be assigned to aromatic =C--H
out-of-plane bending vibrations (see Fig.\,\ref{carbIR}). The
11.3\,$\mu$m feature can be assigned to a single =C--H band, whereas
the band at around 12.3\,$\mu$m is caused by 2--3 and the feature at
13.27\,$\mu$m by 3--4 adjacent H atoms in the periphery of PAH
molecules. The 13.56 and 14.3\,$\mu$m bands can be attributed to around 4
adjacent H atoms bound to the aromatic ring. The observed IR
features in carbonaceous condensates are typical for both
non-substituted PAHs or PAHs containing methyl or methylene groups,
as well as for carbonaceous grains containing aromatic and aliphatic
structures.

Sometimes, also oxygen-containing functional groups such as --C=O
(carbonyl group) are incorporated into the carbonaceous structures
which can be identified at wavelengths between 5.8--6.0\,$\mu$m. In
addition, higher fullerenes can show vibrational modes in this range
(Mordkovich 2000).

Carbonaceous condensates consisting of grain/PAH blends show
spectral characteristics that are comparable to the observed IR
spectra of post AGB stars and protoplanetary nebulae (Kwok et al.\
2001, Hony et al.\ 2003, Hrivnak et al.\ 2007). A comparison shown
in Fig.\,\ref{carbIR} demonstrates that nearly all of the IR bands
observed in the spectrum of the post-AGB-star HD 56126 are also
present in the spectrum of the laboratory low-temperature
condensate. Differences in band ratios, apparent for the =C-H
out-of-plane bands between 11 and 14\,$\mu$m, are due to a higher
abundance of small PAHs, or larger PAHs containing outer rings with
up to 3 to 4 H atoms compared to larger, compactly condensed species
which are thought to be very stable under astrophysical conditions.
Carbonaceous particles built up by large PAHs in the growth process
and purified from the adsorbed, smaller PAHs by extraction show a
better coincidence with the observed bands (see gray solid curve in Fig.\,\ref{carbIR}).
However, in order to obtain a dominance of the 11.3\,$\mu$m
out-of-plane band due to solo =C--H frequently observed in PPNs and
PNs, very large (N$_C$ $\geq$ 100) and rather elongated PAH molecules are
required (\cite{Hony:01}).  The comparison
reveals that the low-temperature condensate is a promising dust
analog for carbonaceous materials produced in carbon-rich AGB stars.

\section{Far-Infrared Spectroscopy}

\subsection{Selected FIR bands of solids: carbonates, phyllosilicates and the 69micron forsterite band}

\begin{figure}
\vspace*{-0.8 cm}
\begin{center}
 \includegraphics[width=3.4in]{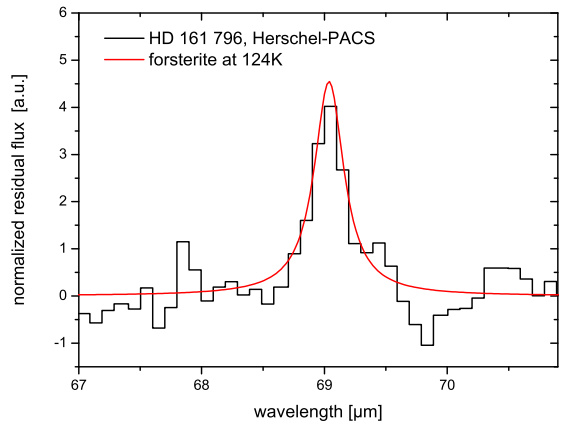}
 \vspace*{-0.5 cm}
 \caption{The 69\,$\mu$m band in the (binned) Herschel-PACS spectrum
 of the post-AGB star HD161796. The 'smooth' line shows the best fit,
 obtained with forsterite at T=124\,K.}
   \label{69my}
\end{center}
\end{figure}

A number of crystalline dielectric solids show phonon bands in the
far-infrared range ($\lambda$$>$50~$\mu$m), where the relatively
cold dust located at larger distances from stars (e.g.\ in planetary
nebulae or protoplanetary systems) is emitting thermal
radiation. If observed, these bands would be of high diagnostic
value to assess the mineralogy of this dust and to draw conclusions
about its history. The profiles of these bands have been
demonstrated to be quite sensitive to temperature, i.e.\ the bands
get sharper and shift to shorter wavelengths at low temperatures.
Consequently, it is necessary to perform laboratory spectroscopy at
these low temperatures for providing comparison spectra, or for the
determination of low-temperature optical constants as input for
calculations, respectively (see Sect.\ 2.2 for the analogous case of
high temperatures).

Such low-temperature spectroscopic data have been produced by
laboratory work in the past years. E.g., temperature-dependent
optical constants have been measured for forsterite (Suto et al.\
2006) and for carbonates (Posch et al.\ 2007). These optical
constants are very valuable because of their flexibility and
independence from the measurement conditions (see Sect.\,2.1). Other
studies have determined the changes of band positions and band
widths with temperature in absorption measurements based on the
pellet technique. E.g., the long-wavelength bands of several
crystalline silicates have already been measured at low temperatures
about ten years ago (\cite{Bowey:02,Chihara:02}).

A more complete investigation of the low-temperature absorption
spectra of olivine minerals with different iron contents was
presented by \cite{Koike:06}. The quantification of the nonlinear
peak shift of the 69\,$\mu$m band and of the band broadening with
temperature allows to use these bands as thermometers under certain
conditions such as at known iron content of the silicate. As
an example of how temperature dependent optical constants can be
used, Fig.\ \ref{69my} shows the 69\,$\mu$m band in the
Herschel/PACS spectrum of the post-AGB star HD161796 (obtained by
the MESS consortium, Groenewegen et al.\ 2011; see also de Vries et
al.\ 2011). A one-temperature fit of the 69\,$\mu$m band indicates
that the forsterite grains in this system have a temperature of the
order of 124 K. For our derivation of the temperature, we used the
laboratory measurements of forsterite's optical constants at
different temperatures by Suto et al.\ (2006).

\begin{figure}
\vspace*{-0.9 cm}
\begin{center}
 \includegraphics[width=2.9in]{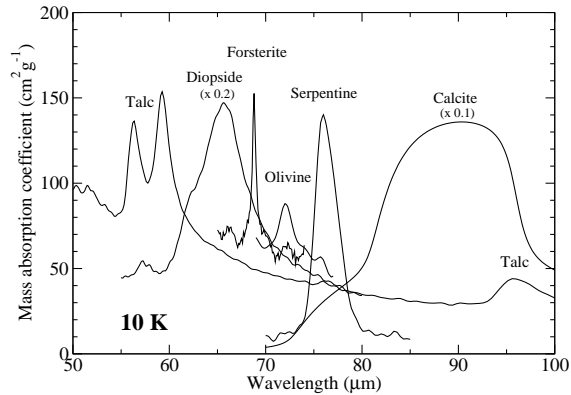}
 \caption{Far-infrared bands of a variety of minerals measured at a
 temperature of 10 K. The calcite spectrum
has been calculated from the optical constants by using the CDE model,
whereas all other spectra have been measured on submicron-sized particles
embedded in polyethylene pellets.}
   \label{fig1}
\end{center}
\end{figure}

Apart from olivines, several other crystalline silicates show clear
FIR bands in the long-wavelength range as well. In Fig.\ \ref{fig1},
bands of diopside (CaMgSiO$_3$) and of the hydrous silicates
serpentine ((Mg,Fe)$_3$Si$_2$O$_5$(OH)$_4$) and talc
(Mg$_3$[Si$_4$O$_{10}$|(OH)$_2$]) are shown for example. Diopside
has a very strong 66\,$\mu$m band (note the five-times reduction of
the peak intensity in our plot). It is an abundant constituent of
early condensates in the solar nebula (see, e.g.\ Posch et al.\
2007a), but the band has not yet been identified in the spectra of
protoplanetary disks. By contrast, it may in fact contribute to a
band seen at this wavelength in planetary nebulae
(\cite{Kemper:02,Chihara:07}).

\begin{figure}
\vspace*{-0.3 cm}
\begin{center}
 \includegraphics[width=3.7in]{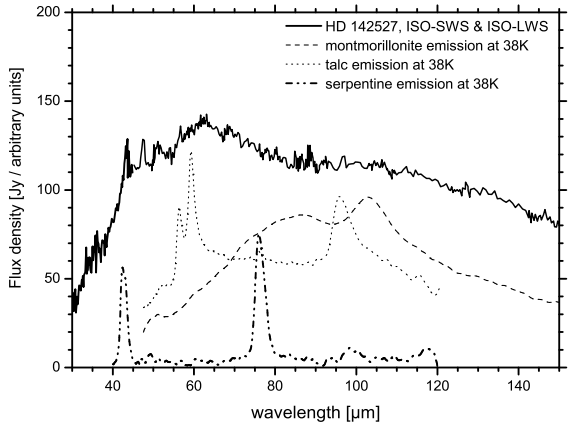}
\vspace*{-0.2 cm}
 \caption{Comparison of simulated emission of cold hydrous silicate dust
with the thermal emission of the Herbig AeBe object HD 142527
(after Mutschke et al.\ 2008).}
   \label{fig2}
\end{center}
\end{figure}

Hydrous silicates are weathering products of silicate minerals
including olivines and pyroxenes and they are abundant in some classes of
primitive meteorites. Hydrous silicates were also discussed to be a
carrier of a broad 100~$\mu$m emission band detected in Herbig Ae/Be
stars (Malfait et al.\ 1999). Far-infrared absorption spectra of
hydrous silicates have been measured by different laboratory
astrophysics groups (Koike et al.\ 1982, Koike \& Shibai 1990,
Hofmeister \& Bowey 2006, Mutschke et al.\ 2008). The spectra show
very interesting features such as the longest-wavelength phonon
bands known to our knowledge (e.g. one at about 270~$\mu$m for
minerals of the chlorite group). However, far-infrared emission features could not be successfully assigned to bands of hydrous silicates so far
(see Fig.\,\ref{fig2}).

Finally, broad far-infrared bands observed in the emission spectrum
of the planetary nebula NGC 6302 have been proposed to be caused by
carbonate dust (Kemper et al.\ 2002). Carbonates have extremely strong
far-infrared bands (note the ten-times reduction of the peak intensity
in Fig.\,\ref{fig1}), so that already small amounts of these minerals,
which may be formed by non-equilibrium reactions of silicates with a
H$_2$O-CO$_2$-rich gas (Toppani et al.\ 2005), should be detectable.
Low-temperature measurements of the optical constants of carbonates
by \cite{Posch:07b} largely confirmed this possible assignment for
NGC 6302, but did not give support to an assignment to a 100~$\mu$m
emission feature observed in protostars (Chiavassa et al.\ 2005).


\subsection{Dust continuum opacities in the submillimeter wavelength range}

\subsubsection{Wavelength and temperature dependence}

Apart from the far-infrared phonon bands, which may provide information
about the composition and temperature of cold, mostly crystalline
dust, the underlying continuum dust opacity is of great importance.
For example, one of the big uncertainties in determining the amount 
of cold dust in interstellar space arises indeed from the
uncertainty in the emissivity of the dust particles at these wavelengths.
Furthermore, models of the spatial dust distribution of circumstellar dust,
e.g.\ in planetary systems (debris disks) depend on the wavelength and
temperature dependence of the dust opacity.

\begin{figure}
\vspace*{-0.2 cm}
\begin{center}
 \includegraphics[width=3.5in]{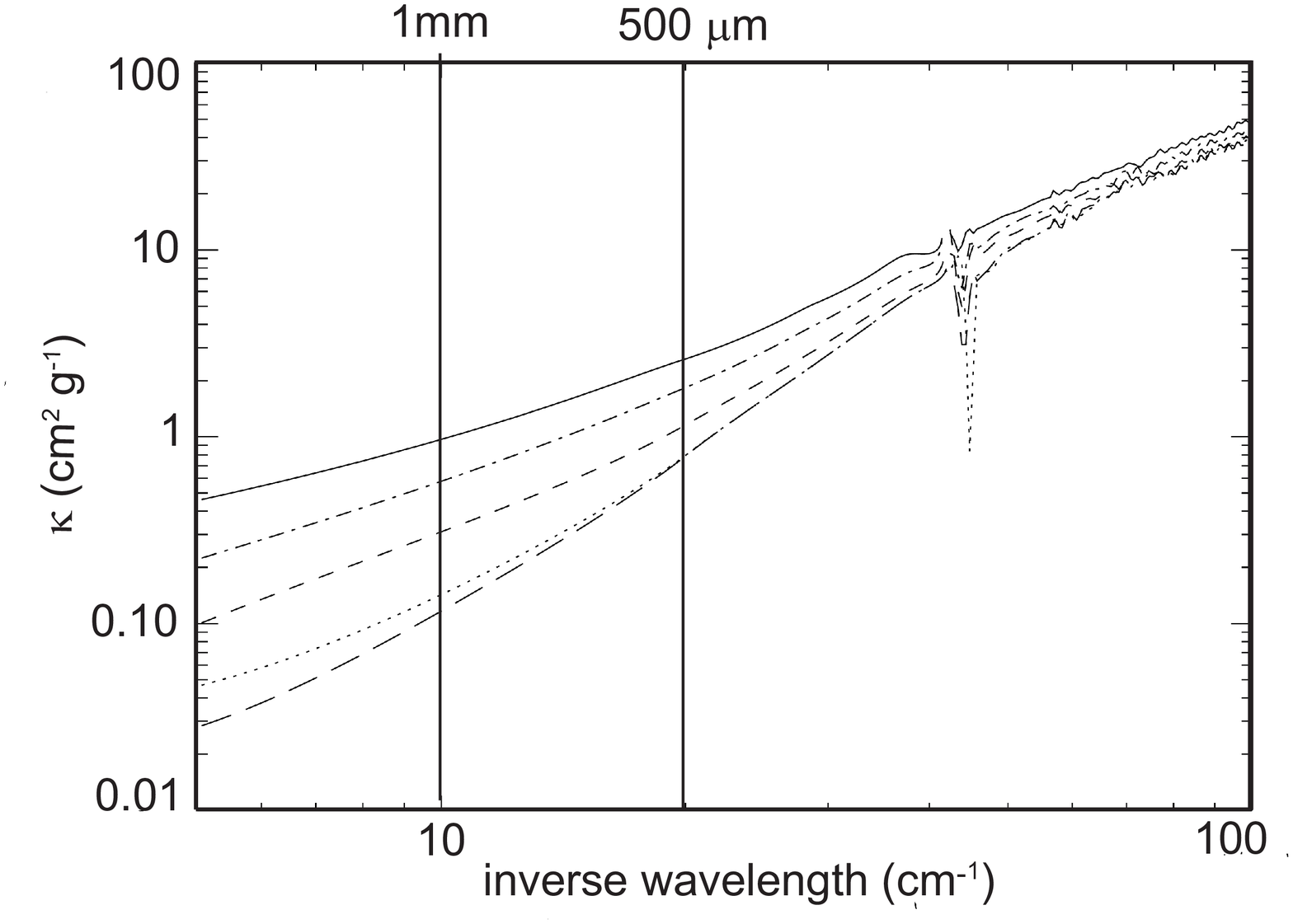}
\vspace*{-0.6 cm}
 \caption{Mass absorption coefficient for MgSiO$_3$, synthetized by sol-gel
reaction at 300\,K (solid line), 200\,K (dot-dashed line), 100\,K (short dashed
line), 30\,K (dotted line), and 10\,K (long-dashed line) (Boudet et al.\ 2005).}
   \label{beta}
\end{center}
\end{figure}

In contrast to the far-infrared bands, the continuum opacity of dust
in most astrophysical environments is determined by that of
amorphous solids, especially amorphous silicates, because their
emissivity is generally higher at far-infrared wavelengths. The
reason for that is the possibility of additional low-energetic
transitions due to structural disorder. Among them, the
disorder-induced phonon transitions give a temperature-independent
contribution which has been shown by \cite{Schloe:64} to lead to a
$\lambda^{-2}$ wavelength dependence at shorter and a $\lambda^{-4}$
dependence at longer wavelengths (see also Meny et al.\ 2007). A
strongly temperature-dependent behavior is, however, introduced by
additional low-energetic tunneling transitions of structural groups
in the amorphous network.

First studies of the temperature-dependent absorption coefficient of
amorphous silicates in the sub-millimeter and millimeter continuum
have been performed by \cite{Agladze:96,HM97} and
\cite{Mennella:98}. Recently, these studies have been extended to a
larger set of amorphous silicate materials measured in a wide
wavelengths and temperature range (Boudet et al.\ 2005, Coupeaud et al.\
2010). In these studies, the power-law
spectral index $\beta$ of absorption was found to increase with
decreasing temperature, especially at wavelengths larger than about
300~$\mu$m. For amorphous Mg silicates and SiO$_2$, this
anti-correlation of T and $\beta$ is demonstrated in
Fig.\,\ref{beta} (values taken from Boudet et al.\ 2005). The spectral
index varies between 1 and 3 for temperatures between 300 and 10\,K,
which confirms the expectations from theory.

In Tab.\,\ref{tab2} we summarize opacity values and power-law indices for a
number of amorphous and crystalline Mg- and Fe-silicates, taken from four
of the mentioned publications (apologies for being not complete). Additionally,
we compare with values measured for an amorphous carbon by \cite{Mennella:98}
and with the numbers calculated for spherical grains composed of the model
material ``astronomical silicate (astrosil)'' (Li \& Draine 2001).

\begin{table}
  \begin{center}
  \caption{Continuum opacities of dust species at low temperatures (4-24~K).
  The values are given as mass absorption coefficients at 100~$\mu$m
(read from figures -- approximate) and 1\,mm wavelength, plus the
power-law slope value $\beta$ at 1mm wavelength (for forsterite, we
refer to the slope at 40-80~$\mu$m). Room-temperature values are
given in parentheses.}
  \label{tab2}
 {\scriptsize
  \begin{tabular}{|l|c|c|c|c|}\hline
{\bf Dust species} & {\bf $\kappa$ (100~$\mu$m)} & {\bf $\kappa$ (1~mm)} & {\bf $\beta$} & {\bf Reference} \\
& {\bf [cm$^2$g$^{-1}$]} & {\bf [cm$^2$g$^{-1}$]} & & \\ \hline
Am.\ Fe-silicate (FAYA)& $ 95 (110)$ & $ 0.86 (5.0)$ & $2.04 (1.35)$ & Mennella et al.\ (1998) \\
Am.\ Mg-Silicate (Mg SiO$_3$ glass)& $ 45 (50)$ & $ 0.22 (0.75)$ & $2.14 (1.58)$ & Boudet et al.\ (2005) \\
Am.\ Mg-Silicate (Mg SiO$_3$ sol-gel)& $ 40 (50)$ & $ 0.12 (0.98)$ & $2.74 (1.44)$ & Boudet et al.\ (2005) \\
Am.\ Mg-Silicate (Mg SiO$_3$ sol-gel)&  & $ 0.28 $ & $ \sim2.0 $ & Agladze
et al.\ (1996) \\
``Astronomical silicate'' (spheres)& $ 29 $ & $ 0.30 $ & $1.6$ & Li \& Draine (2001) \\
Forsterite & $  2.1 (20) $ &  & $5.0 (2.6)$ & Chihara et al.\ (2001) \\
Forsterite & $  80 (120) $ & $0.16 (0.43)$ & $2.32 (2.04)$ & Mennella et al.\ (1998)\\
Forsterite (unheated) &  & $0.22$ & & Agladze et al.\ (1996) \\
Am. Carbon (BE) & $300 (500)$ & $21 (72)$ & $1.15 (0.76)$ & Mennella et al.\ (1998) \\
\hline
  \end{tabular}}
 \end{center}
\end{table}

At a wavelength of 1~mm, the low-temperature opacity values of
amorphous Mg silicates seem to converge to values of
0.1--0.3~cm$^2$g$^{-1}$, which is close to the ``astrosil'' value.
However, we have to mention that Agladze et al.\ (1996) have reported
higher values up to 3.7~cm$^2$g$^{-1}$ for some of their sol-gel
produced silicate samples as well. The reason for that is not clear,
but may have to do with a high porosity of these samples.
Fe-silicates seem to have somewhat higher opacities, and a somewhat
stronger temperature dependence, but this is only based on one
measurement so far (Mennella et al.\ 1998). For mixed Mg/Fe
silicates, investigations are still lacking. The power-law exponent
around 1\,mm  wavelength has typically values between 2 and 3 for
amorphous silicates. The ``astrosil'' value has been reduced in
order to better match galactic dust emission data. Such values seem
to correspond to certain data in the Agladze 
et al.\ measurements, but just these have unusually 
high opacity values at the same time.

For amorphous carbon, a similar temperature dependence of the
opacity as for amorphous silicates has been reported by Mennella et
al.\ (1998), although at a strongly enhanced opacity level (see
Tab.\,\ref{tab2}). The excitation mechanisms behind this could
actually be totally different because contributions from free charge
carriers might be dominating (see following subsection).

For crystalline silicates, the measured values do not seem to be in perfect agreement so far. For forsterite, Mennella et al.\ (1998) and
Agladze et al.\ (1996) found low-temperature opacity and
$\beta$ values, which were similar to those of the amorphous
silicates, although for the same composition $\kappa$ values lower
by a factor of four were reported (Agladze et al.\ 1996) and the
temperature dependence was considerably weaker. Chihara et al.
(2001), however, measured values at 100~$\mu$m wavelengths, which
were only one order of magnitude higher, and therefore in
disagreement with a $\beta\sim$\,2 behavior. This discrepancy
certainly deserves further investigation.

\subsubsection{Grain shape dependence and carbonaceous materials}

For carbonaceous materials, the FIR-behavior is not only influenced
by the internal structure, but also strongly by the morphology of
the carbon grains. It is well known that the clustering of carbon
grains decreases the spectral index $\beta$ very strongly
(\cite{Stog:95,Mich:96}). The influence of the internal structure on
the far-infrared behavior of differently structured carbon materials
was already studied by \cite{Koi:95}, Mennella et al.\ (1998), and
\cite{Pap:96}. These measurements were performed in transmission
using polyethylene pellets. However, one has to keep in mind that
this method cannot disentangle the  structural and morphological
effects on the FIR absorption.

\begin{figure}
\vspace*{-0.4 cm}
\begin{center}
 \includegraphics[width=4.3in]{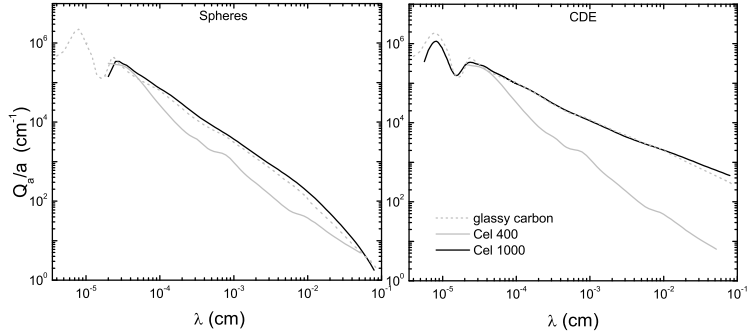}
 \vspace*{-0.3 cm}
 \caption{Absorption efficiency divided by particle radius calculated
for spherical particles and ellipsoidal grains in vacuum from the
optical data of pyrolized cellulose materials.}
   \label{carbFIR}
\end{center}
\end{figure}

For the interpretation of observational FIR data provided by
ground-based or satellite-borne observations such as with the
Herschel satellite, the opacity of small grains in vacuum is
required, which can be calculated from the optical constants $n$ and $k$
for spherical shapes by the Mie theory. In addition, shape effects
can roughly be simulated by a continuous distribution of ellipsoids
with the same probability of all shape parameters (CDE) in the
Rayleigh limit (Bohren \& Huffman 1983). In Fig.\,\ref{carbFIR}, the calculated
normalized absorption efficiencies of two carbon materials ('cel 400'
and 'cel 1000') are compared with those of glassy carbon. The two
materials represent a sp$^3$ hybridization dominated non-conducting
material ('cel 400') and an aromatic carbonaceous material with a high
number of free charge carriers ('cel 1000'; see J\"ager et al.\ 1998).

Whereas in case of the aliphatic-dominated carbon material 'cel 400'
the spectral index $\beta$ in the FIR region ($\lambda$ $\ge$
100~$\mu$m) is about 1.25 irrespective of the particle model, for
the aromatic sample carbonized at 1000\,$^{\rm o}$C it is 2.28 for
spherical particles, but only 0.71 for agglomerated (CDE) particles.
The latter value is very close to the one measured by Mennella et
al.\ (1998) for amorphous carbon particles embedded in a polyethylene
pellet (see Table 1), although the absolute absorption coefficient
is in between those calculated for spherical and CDE particles. This
may indicate that in the pellet the carbon particles are strongly
aggregated. We conclude that for cosmic particles formed of conducting
materials such as highly graphitized amorphous carbon, the far-infrared emissivity can be very high, but this will strongly depend on the grain morphology.


\vspace{1.0cm}


{\bf Discussion}:\\

\textsc{Goumans:}  You mentioned that if you have the optical constants,
one can get the spectrum of grains of any shape and size.
What about very small grains, or clusters, that do not behave like
bulk material any more? Is spectroscopy on those desirable
and possible?

\textsc{Posch:}  Studies of clusters, including infrared
spectroscopy, can certainly help to clarify the nature of some
unidentified broad infrared bands that were hitherto assigned to
mineral species. However, IR spectroscopy of clusters requires very
sophisticated techniques and will therefore take even more time than
solid state spectroscopy. Additionally, the spectra critically
depend on the cluster size, so experiments must be carried out for a
wide range of possible cluster sizes.


\end{document}